\documentclass[conference]{IEEEtran}
\usepackage{amsmath,amsthm}
\usepackage{cite}
\usepackage{graphicx}
\usepackage{epstopdf}
\usepackage{amsfonts,amsmath,amssymb}
\usepackage{cite}
\usepackage{graphicx}
\usepackage{url}
\usepackage{bm}
\usepackage{bbm}
\usepackage{amssymb}

\begin{document}

\title{{Gaussian Entanglement Distribution via Satellite}}
\author{
\IEEEauthorblockN{Nedasadat Hosseinidehaj, Robert Malaney}
\IEEEauthorblockA{School of Electrical Engineering  \& Telecommunications,\\
The University of New South Wales,\\
Sydney, NSW 2052, Australia\\
neda.hosseini@student.unsw.edu.au, r.malaney@unsw.edu.au}
}

\vspace{-5cm}

\maketitle
\begin{abstract}
In this work we analyze  three quantum communication schemes  for the  generation of Gaussian entanglement between two ground stations. Communication occurs via a satellite over two independent atmospheric fading channels dominated by turbulence-induced beam wander. In our first scheme the engineering complexity remains largely on the ground transceivers, with the satellite acting simply as a reflector. Although the channel state information of the two atmospheric channels remains unknown in this scheme,  the Gaussian entanglement generation between the ground stations can still be determined. On the ground, distillation and Gaussification procedures can be applied, leading to a refined Gaussian entanglement generation rate between the ground stations. We compare the rates produced by this first scheme with two competing schemes in which quantum complexity is added to the satellite, thereby illustrating the trade-off between space-based engineering complexity  and the rate of ground-station entanglement generation.

\end{abstract}

\section{Introduction}

Although current quantum communication systems are  limited to relatively small scales, it is widely anticipated that  next-generation
quantum networks will in some capacity invoke the concept of free-space optical (FSO) communications (for review see \cite{fso}) in order to extend the communication range. Coupled to this, is the growing belief that
space-borne quantum transceivers will soon make full blown \emph{global} quantum communications an engineering reality \cite{r7,r8,r9,r10,r11,r12,r13,r14,r15,r16,r17}. While it remains to be seen whether FSO quantum communications  will be dominated by discrete single photon (qubit) technology, multi-photon continuous variable (CV) technology, or even some hybrid of both technologies\cite{lockbook}, it is important to fully understand the capabilities of both types of technologies in the free-space channel.

 Previous work in the satellite quantum communication scenario  has largely focussed on qubit technologies. In this work we will focus on the CV scenario, with the aim of assessing some of the different CV quantum-communication architectures that could be  deployed through atmospheric  fading channels. Our specific interest will be on the distribution of Gaussian quantum entanglement over the different architectures. Gaussian entanglement between quantum states has been widely recognized as a basic resource for quantum information processing and quantum communications (for review see \cite{A,rr1,rr2,1}). Here, we will analyze three different space-based schemes  that will allow for Gaussian entangled states to be shared at separate ground stations.

  In our first scheme, referred to as \emph{direct transmission entanglement}, no quantum technology is deployed at the satellite - the satellite is utilized simply in a reflector mode \cite{foot1}.
 The main motivation for this is that quantum engineering is a highly sophisticated business, demanding leading-edge technology. Having such technology based in hard-to-reach satellite systems could potentially make global quantum communication systems less reliable (due to the rarity of maintenance), and costly to update as new  quantum technology matures. Relatively speaking, one could consider a reflection at the satellite as a low space-based-complexity system. In this system a two-mode entangled squeezed state  is generated at ground station A, with one component of the beam held at A and the other component transmitted to ground station B via a low-earth-orbit (LEO) reflecting relay satellite. As a proof of concept on the reflecting paradigm, we note  the recent experimental tests of  \cite{r16,r17} in which photons  were reflected (and subsequently detected) off a LEO satellite.

\begin{figure}[!t]
    \begin{center}
   {\includegraphics[width=2 in, height=4 in]{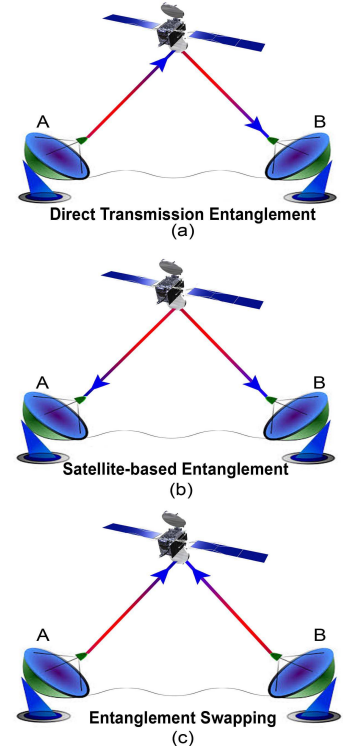}}
    \caption{(Color online) Entanglement generation schemes. (a) is the direct transmission scheme where reflection is used at the satellite. (b) is a  scheme where entangled photon generation takes place directly in the satellite and then distributed in separate downlinks to the ground stations. (c) is a  scheme in which the Gaussian entangled states are transmitted independently through two fading channels from two ground stations to the satellite. They are then swapped via a Bell-measurement at the satellite, resulting in creation of a new entanglement between the ground stations.
    }\label{fig:both}
    \end{center}
\end{figure}

 The other schemes we study can be considered as high space-based-complexity in that they involve the deployment of quantum technology at the satellite. Our second scheme, referred to as \emph{satellite-based entanglement}, invokes entanglement generation in the satellite itself with subsequent transfer to the ground stations directly. Our third scheme, referred to as \emph{entanglement swapping}, utilizes on-board Gaussian entanglement swapping between arriving beams of photons entangled with (and emitted by) separate ground stations. All three of our schemes are illustrated in Fig.~\ref{fig:both}.

 In all schemes, the  transmitted beam  will encounter fluctuations (fading) caused by its traversal (twice) through the atmosphere. Among the many unwanted disturbances in realistic atmospheric channels, we will concentrate here on transmission fluctuations caused by beam wander, an effect anticipated to dominate the noise contributions  in many scenarios \cite{fso,20,21}.

 It is the first aim of this work to provide a quantitative assessment, in terms of resulting Gaussian entanglement, of the low space-based-complexity scheme in
 relation to the two high space-based-complexity  schemes. A second aim of our work is to explore post-processing strategies that can occur at the receiving ground stations. Due to the fluctuating fading channels traversed by the beams, a non-Gaussian mixed state is produced.  At the receiver a post-selection strategy can be deployed  in order to distill (concentrate) the Gaussian entanglement between the two ground stations. Such post-selection strategies  could be based on quantum measurement techniques, or on classical measurements of the channel transmittance. However, such  classical measurements of the  channel transmittance will require additional complexity in the transmission/detection strategy.  We will be specifically interested in investigating the gain in Gaussian entanglement obtained by the inclusion of this additional classical complexity.

 Note that, although  largely motivated  by the  use of FSO in the scenario of satellite quantum communications to/from terrestrial ground stations \cite{r7,r8,r9,r10,r11,r12,r13,r14,r15,r16,r17}, our results are applicable to a range of FSO links such as high altitude platform (HAP)-to-satellite quantum links  \cite{hap} and aircraft-to-ground quantum links \cite{aircraft}.
 The range of atmospheric channels we study will cover all of these different quantum communication scenarios.

The structure of the remainder of this paper is as follows. In Section II, the basic concepts of Gaussian quantum states are introduced, and our three transmission schemes over terrestrial-satellite fading channels are analyzed in terms of output covariance matrices. In Section III, the performances of the three schemes are compared in terms of output Gaussian entanglement. In Section IV, we discuss the impact of the post-selection strategy utilized at the ground stations. Finally, concluding remarks and future research directions are provided in Section V.

%

\section{ Quantum communication over fading channels}

 In the following we discuss the three quantum communication schemes of  Fig.~\ref{fig:both}, but first introduce some features that will be needed for their description.


For a single bosonic mode with annihilation and creation operators $\hat a,\,{\hat a^\dag }$, the quadrature operators $\hat q,\hat p$ are defined by
$\hat q = \hat a + \,{\hat a^\dag }\,,\,\,\,\,\,\hat p = i({\hat a^\dag } - \hat a\,)$
 which satisfy the commutation relation $\left[ {\hat q,\,\hat p} \right] = 2i$ (here $\hbar=2$).
 The vector of quadrature operators for a quantum state with $n$ modes can be defined as
${\hat R_{1, \ldots ,n}} = \left( {{{\hat q}_1},\,{{\hat p}_1}, \ldots ,{{\hat q}_n},{{\hat p}_n}\,} \right)$.
Similarly, ${R_{1, \ldots ,n}} = \left( {{q_1},\,{p_1}, \ldots ,{q_n},{p_n}\,} \right)$ is defined for the real variables $q, p$ - the eigenvalues of the quadrature operators.

We will discuss both non-Gaussian and Gaussian states, the latter being states  whose Wigner function is a Gaussian distribution of the quadrature variables. Gaussian states are completely characterized by the first moment of the quadrature operators $\left\langle {{{\hat R}_{1, \ldots ,n}}} \right\rangle $ and a \emph{covariance matrix} (CM) $M$, \emph{i.e.} a matrix of the second moments of the quadrature operators as
\begin{eqnarray}\label{ad1}
{M_{ij}} = \frac{1}{2}\left\langle {{{\hat R}_i}{{\hat R}_j} + {{\hat R}_j}{{\hat R}_i}} \right\rangle  - \left\langle {{{\hat R}_i}} \right\rangle \left\langle {{{\hat R}_j}} \right\rangle .
\end{eqnarray}
The CM of an $n$-mode quantum state is a $2n \times 2n$ real and symmetric matrix which must satisfy the uncertainty principle, \emph{viz.},
$M  + i\,\Omega \, \ge \,0$,
where
\begin{eqnarray}\label{ad2}
\Omega : = \mathop  \oplus \limits_{k = 1}^n \,\,\omega  = \left( {\begin{array}{*{20}{c}}
\omega &{}&{}\\
{}& \ddots &{}\\
{}&{}&\omega
\end{array}} \right)\,,\,\,\omega : = \left( {\begin{array}{*{20}{c}}
0&1\\
{ - 1}&0
\end{array}} \right).
\end{eqnarray}
By local unitary operators, the first moment of every two-mode Gaussian state can be set to zero and the CM can be transformed into the following standard form
\begin{eqnarray}\label{p3}
{M_{s}}= \left( {\begin{array}{*{20}{c}}
A&C\\
{{C^T}}&B
\end{array}} \right),\,
\end{eqnarray}
where
$A = aI\,,\,B = bI\,,\,C = diag\left ( {{c_ + },{c_ - }} \right )$,
 $a,b,{c_ + },{c_ - } \in \mathbb{R}$, and
 $I$ is a $2 \times 2$ identity matrix.
Considering the standard form, the symplectic spectrum of a partially transposed CM is given by
\begin{eqnarray}\label{p5}
{\nu _ \pm } = \sqrt {\frac{{\Delta  \pm \sqrt {{\Delta ^2} - 4\det M_s } }}{2}} ,
\end{eqnarray}
where $\Delta = \det A + \det B - 2\det C$. A quantitative measure of Gaussian entanglement can be derived in terms of the logarithmic negativity ${E_{_{LN}}}\left( {{M_s}} \right) = \max \left[ {0, - {{\log }_2}\left( {{\nu _ - }} \right)} \right]$, where ${\nu _ - }$, as given above, is the smallest symplectic eigenvalue of the partially transposed CM \cite{A}.



In free-space channels the transmittance  fluctuates due to atmospheric effects. Such fading channels can be
 characterized by a distribution of transmission coefficients $\eta$ with a probability density distribution $p(\eta)$. The main contributors to transmission losses in free-space quantum communication are atmospheric turbulence, diffraction, scattering, and absorption. Diffraction, scattering, and absorption are all wavelength dependent, and to a large extent can  be mitigated by an appropriate choice of communication wavelength \cite{fso, r10}. Atmospheric turbulence arises due to random fluctuations in the refractive index caused by stochastic temperature variations. This effect leads to beam wandering  as well as beam broadening \cite{fso, r10}. In this paper, we take the usual assumption  that  transmittance fading is  largely dominated by beam wander \cite{fso,20,21,foot2}. Beam wandering causes the beam center to be randomly displaced from the aperture center in the receiver plane. Assuming that the beam-center position is normally distributed with variance $\sigma _b^2$ around a point at a distance of $d$ from the aperture center, the beam-deflection distance fluctuates according to the Rice distribution, which results in the probability density distribution $p\left( \eta  \right)$ being given by the log-negative generalized Rice distribution \cite{20}. Unlike earlier models, e.g. the log-normal distribution, the log-negative generalized Rice distribution more accurately describes the operationally-important  transmission distribution tail \cite{20}. In the particular case, $d = 0$,
when the beam spatially fluctuates around the center of the receiver's aperture  such fading can be described by the log-negative Weibull distribution \cite{20} \cite{21},
\begin{eqnarray}\label{f1}
p\left( \eta  \right) = \frac{{2{L^2}}}{{\sigma _b^2\lambda \eta }}{\left( {2\ln \frac{{{\eta _0}}}{\eta }} \right)^{\left( {\frac{2}{\lambda }} \right) - 1}}\exp \left( { - \frac{{{L^2}}}{{2\sigma _b^2}}{{\left( {2\ln \frac{{{\eta _0}}}{\eta }} \right)}^{\left( {\frac{2}{\lambda }} \right)}}} \right)
\end{eqnarray}
for $\eta  \in \left[ {0,\,{\eta _0}} \right]$, with $p\left( \eta  \right) = 0$ otherwise.
Here, $\sigma _b^2$ is the beam wander variance,
 $\lambda$ is the shape parameter,  $L$ is the scale parameter, and ${\eta _0}$ is the  maximum transmission value. The latter three parameters are given by
 \begin{eqnarray}\label{f2}
\begin{array}{l}
\lambda  = 8h\frac{{\exp \left( { - 4h} \right){I_1}\left[ {4h} \right]}}{{1 - \exp \left( { - 4h} \right){I_0}\left[ {4h} \right]}}{\left[ {\ln \left( {\frac{{2\eta _0^2}}{{1 - \exp \left( { - 4h} \right){I_0}\left[ {4h} \right]}}} \right)} \right]^{ - 1}}\\
\\
L = \beta{\left[ {\ln \left( {\frac{{2\eta _0^2}}{{1 - \exp \left( { - 4h} \right){I_0}\left[ {4h} \right]}}} \right)} \right]^{ - \left( {{1 \mathord{\left/
 {\vphantom {1 \lambda }} \right.
 \kern-\nulldelimiterspace} \lambda }} \right)}}\\
\\
\eta _0^2 = 1 - \exp \left( { - 2h} \right) ,
\end{array}
\end{eqnarray}
where ${I_0}\left[ . \right]$ and ${I_1}\left[ . \right]$ are the modified Bessel functions, and where $h = {\left( {{\beta \mathord{\left/
 {\vphantom {a W}} \right.
 \kern-\nulldelimiterspace} W}} \right)^2}$, with $\beta$ being the aperture radius and  $W$  the beam-spot radius.

Note that, assuming fixed values for $W$ and $\beta$, the transmittance mean value $\left\langle \eta  \right\rangle $ always decreases with increasing ${\sigma _b}$. Also note, that the uplink (ground-to-satellite) first traverses the atmosphere followed by a larger-scale free-space traversal, whereas the downlink (satellite-to-ground) does the opposite. For
the case of fixed fading parameters $W$ and $\beta$,  this means that in general the beam wander variance $\sigma _b^2$ for the uplink is significantly larger than the downlink \cite{fso}. Finally, note that the rate of  atmospheric fluctuations we consider are of order kHz, which is at least a thousand times slower than typical transmission/detection rates \cite{fso,21}. This means that channel measurements can be obtained at the cost of additional (classical) transmission/receiver complexity. We will assume that such measurements are in place at the ground receivers (only) in our first two schemes. As we shall see, in order to optimize our third scheme (entanglement swapping) we will need the additional complexity of classical channel measurements at the satellite. Channel measurements could be made via several schemes - \emph{e.g.,} via coherent (classical) light pulses that are intertwined with the  quantum information or via the traversal through the atmosphere of a local oscillator \cite{sem}. We will explore later the cost (in terms of Gaussian entanglement) of removing this additional classical complexity.



\subsection{Direct Transmission Entanglement}
The direct transmission scheme illustrated in Fig.~\ref{fig:both}(a) is now analyzed and the CM of the output state calculated. Let us consider the ground station A initially possessing a Gaussian two-mode entangled squeezed state. We assume one mode remains at the ground station while the other mode is transmitted over the fading uplink to the satellite, then perfectly reflected in the satellite and sent through the fading downlink toward the ground station B. As a result, depending on the initial degree of entanglement, there would exist an entangled state between the two ground stations. Note that  we assume the separate uplink and downlink channels  are independent and non-identical.

Now let us assume that the initial entangled states are two-mode squeezed vacuum states with squeezing $r$, then their initial CM
 can be written
\begin{eqnarray}\label{D1}
{{M}_i} = \left( {\begin{array}{*{20}{c}}
{v\,I}&{\sqrt {{v^2} - 1} \,Z}\\
{\sqrt {{v^2} - 1} \,Z}&{v\,I}
\end{array}} \right) ,
\end{eqnarray}
where $v = \cosh \left( {2r} \right)$,  $r \in \left[ {0,\,\infty } \right)$, and $Z = diag\left( {1, - 1} \right)$.  Note, the initial entangled states can be coherently displaced without changing the above CM.

After transmission of the optical mode through the uplink and then reflection through the downlink with probability density distributions ${p_{AS}}\left( \eta  \right)$ and ${p_{SB}}\left( {\eta } \right)$, respectively, the CM of the two-mode  state at the ground stations for two realization of the transmission factors ${\eta}$  (uplink) and $\eta '$ (downlink) can be constructed. It is straightforward to show that assuming no additional noise sources this CM is given by
\begin{eqnarray}\label{CP4aa}
{M _{\eta \,\eta '}} = \left( {\begin{array}{*{20}{c}}
{v\,I}&{\sqrt {\eta \,\eta '} \sqrt {{v^2} - 1} \,Z}\\
{\sqrt {\eta \,\eta '} \sqrt {{v^2} - 1} \,Z}&{\left( {1 + \eta \,\eta '\left( {v - 1} \right)} \right)\,I}
\end{array}} \right) .
\end{eqnarray}
Therefore, the elements of the final CM of the resulting mixed state are calculated by averaging the elements of ${{M _{\eta \,\eta '}} }$ over all possible transmission factors of the two fading channels giving
\begin{eqnarray}\label{CP5}
\begin{array}{l}
{M} = \left( {\begin{array}{*{20}{c}}
{v\,I}&{{c}\,Z}\\
{{c}\,Z}&{{b}\,I}
\end{array}} \right)\,, \ {\rm where}
\\ \\
{b} = \int_0^{{\eta _0}} {\int_0^{{{\eta '_0}}} {{p_{AS}}(\eta )\,{p_{SB}}(\eta ')} } \,\left( {1 + \eta \,\eta '\left( {v - 1} \right)} \right)\,d\eta \,d\eta '\\
\\
{c} = \int_0^{{\eta _0}} {\int_0^{{{\eta '_0}}} {{p_{AS}}(\eta )\,{p_{SB}}(\eta ')} } \,\sqrt {\eta \,\eta '} \sqrt {{v^2} - 1} \,d\eta \,d\eta ' .
\end{array}
\end{eqnarray}
Note that since ${\eta}$ and $\eta '$ are  random variables, the  final state ensemble is a non-Gaussian mixture of the Gaussian states obtained for each realization of ${\eta}$ and $\eta '$. Note also, in this scheme it is only the \emph{combined} channel transmissivity ${\eta}\eta '$ that is measured at the ground station B.
\subsection{Satellite-based Entanglement}
In this section the quantum communication scheme in Fig.~\ref{fig:both}(b) is analyzed and CM of the output state between the terrestrial stations is computed. Here a two-mode entangled state is directly generated within the satellite, with both modes then sent over separate fading downlinks to the ground stations. Again we assume that the initial entangled state is a two-mode squeezed  state  described by CM $M_i$ of \eqref{D1}. After distribution of the  modes through the downlink to station A and downlink to station B characterized by probability density distributions ${p_{SA}}\left( \eta  \right)$ and ${p_{SB}}\left( {\eta } \right)$ respectively, the CM of the two-mode Gaussian state between the ground stations for each realization of ${\eta}$ and $\eta '$ is given by
\begin{eqnarray}\label{sa1}
{{M '_{\eta \,\eta '}}} = \left( {\begin{array}{*{20}{c}}
{\left( {1 + \eta \left( {v - 1} \right)} \right)\,\,I}&{\sqrt {\eta \,\eta '} \sqrt {{v^2} - 1} \,Z}\\
{\sqrt {\eta \,\eta '} \sqrt {{v^2} - 1} \,Z}&{\left( {1 + \eta '\,\left( {v - 1} \right)} \right)\,I}
\end{array}} \right) .
\end{eqnarray}
Here, the two fading downlinks are independent and non-identical. The elements of the final CM are simply the average of the elements of ${{M '_{\eta \,\eta '}}}$ over all possible fluctuating transmission factors of the two fading channels giving
\begin{eqnarray}\label{sa2}
\begin{array}{l}
{{M'}} = \left( {\begin{array}{*{20}{c}}
{\,{{a'}}\,I}&{\,{{c'}}\,Z}\\
{{{c'}}\,Z}&{\,{{b'}}\,I}
\end{array}} \right)\ , \ {\rm where} \\
\\
{{a'}} = \int_0^{{\eta _0}} {{p_{SA}}(\eta )} \,\left( {1 + \eta \left( {v - 1} \right)} \right)\,d\eta \,\\
\\
{{b'}} = \int_0^{{{\eta '_0}}} {{p_{SB}}(\eta ')} \,\left( {1 + \eta '\,\left( {v - 1} \right)} \right)\,d\eta '\\
\\
{{c'}} = \int_0^{{\eta _0}} {\int_0^{{{\eta '_0}}} {{p_{SA}}(\eta )\,{p_{SB}}(\eta ')} } \,\sqrt {\eta \,\eta '} \sqrt {{v^2} - 1} \,d\eta \,d\eta ' .
\end{array}
\end{eqnarray} 	
 Again, the  final state ensemble is a non-Gaussian mixture. In this scheme, the individual channel transmissivities ${\eta}$ and $\eta '$ are obtainable via measurements at the ground stations.

\subsection{Entanglement Swapping}
The protocol of entanglement swapping as shown in Fig.~\ref{fig:both}(c) is now analyzed over fading channels, and the CM of the optimal output state computed.
Entanglement swapping \cite{4} is a standard protocol to establish entanglement between distant quantum systems that have never interacted \cite{M, 6, EE}. It is the central mechanism of quantum repeaters \cite{5}, enabling distribution of entanglement over large distances. Previously, the implementations of a swapping-based protocol in the context of CV technology has been studied mostly through fixed attenuation channels \emph{e.g.} \cite{M,19}. In \cite{19}, optimal entanglement swapping with Gaussian states over a lossy optical fiber with fixed attenuation has been analyzed, and we build on this analysis here in the context of two independent fading channels.

  In the entanglement swapping scheme,  each ground station initially possesses a Gaussian two-mode entangled state. One mode of each entangled state is kept by the ground station and the second mode of each state is transmitted to the satellite through a fading uplink. Here, the two fading uplinks are independent and non-identical with probability density distributions ${p_{AS}}\left( \eta  \right)$ and ${p_{BS}}\left( {\eta } \right)$ for the station-A uplink and the station-B uplink, respectively.

Let us consider the  entangled states initially at the ground stations to be a pair of two-mode squeezed vacuum states with the same level of squeezing $r$, with  modes 1 and 2 owned by  ground station A, and modes 3 and 4 owned by  ground station B. These pairs of entangled states will possess CMs described by \eqref{D1}, that is
\begin{eqnarray}\label{CP1}
{M ^{1,2}} = {M ^{3,4}} = \left( {\begin{array}{*{20}{c}}
{v\,I}&{\sqrt {{v^2} - 1} \,Z}\\
{\sqrt {{v^2} - 1} \,Z}&{v\,I}
\end{array}} \right) .
\end{eqnarray}
After transmission of mode 2 through the uplink from station A and transmission of mode 3 through the uplink from station B, prior to any interaction at the satellite the transmitted states  for each realization of ${\eta}$ and $\eta '$ are described by two states with CMs,
\begin{eqnarray}\label{1}
\begin{array}{l}
M_\eta ^{1,2} = \left( {\begin{array}{*{20}{c}}
{v\,I}&{\sqrt \eta  \sqrt {{v^2} - 1} \,Z}\\
{\sqrt \eta  \sqrt {{v^2} - 1} \,Z}&{\left( {1 + \eta (v - 1)} \right)\,I}
\end{array}} \right)\\
\\
M_{\eta '}^{3,4} = \left( {\begin{array}{*{20}{c}}
{\left( {1 + \eta '(v - 1)} \right)\,I}&{\sqrt {\eta '} \sqrt {{v^2} - 1} \,Z}\\
{\sqrt {\eta '} \sqrt {{v^2} - 1} \,Z}&{v\,I}
\end{array}} \right) .
\end{array}
\end{eqnarray}

When the two transmitted modes are received, they are swapped via a Bell measurement at the satellite. First, transmitted modes 2 and 3 are mixed through a balanced beam-splitter, yielding output modes $u$ and $v$. Then, the new quadratures ${\hat q_u}$ and ${\hat p_v}$ are measured by two homodyne detectors, providing the outcomes ${q'_u}$ and ${p'_v}$. 
In order to complete the swapping process, the satellite broadcasts the Bell measurement results so that the two ground stations can properly displace their modes according to the measurement outcomes ${q'_u}$ and ${p'_v}$. In practice, the displacements can be weighted by gain factors to improve the quality of the swapped entanglement. It can be shown (see Appendix A) that if the gains applied to the displacements of modes 1 and 4 are given by
\begin{eqnarray}\label{21}
\begin{array}{l}
{g_1} = \frac{{\sqrt \eta  \sqrt {{v^2} - 1} }}{{2 + \left( {\eta  + \eta '} \right)\left( {v - 1} \right)}} , \ \ \
{g_4} = \frac{{\sqrt {\eta '} \sqrt {{v^2} - 1} }}{{2 + \left( {\eta  + \eta '} \right)\left( {v - 1} \right)}} ,
\end{array}
\end{eqnarray}
 then the CM of the conditional state of modes 1 and 4 (averaged over all possible Bell measurements) at the ground stations is given by
\begin{eqnarray}\label{16s}
\begin{array}{l}
{{M''_{\eta \eta '}}} = \left( {\begin{array}{*{20}{c}}
{\left( {v - \eta m} \right)\,I}&{\sqrt {\eta \,\eta '} m\,Z}\\
{\sqrt {\eta \,\eta '} m\,Z}&{\left( {v - \eta 'm} \right)\,I}
\end{array}} \right) ,  {\rm where}
\\
m = \frac{{\,({v^2} - 1)}}{{2 + \left( {\eta  + \eta '} \right)\left( {v - 1} \right)}} .
\end{array}
\end{eqnarray}

The final (ensemble averaged) swapped state shared by the ground stations is the mixture of the swapped states after each realization of ${\eta}$ and $\eta '$. The total CM of the resulting mixed swapped state is obtained by averaging elements of ${M''_{\eta \eta '}}$ in \eqref{16s} over all possible transmission factors of the two fading channels, giving
\begin{eqnarray}\label{22aa}
\begin{array}{l}
M'' = \left( {\begin{array}{*{20}{c}}
{a''\,I}&{c''\,Z}\\
{c''Z}&{b''\,I}
\end{array}} \right)  , {\rm where}\\
\\
a'' = \int_0^{{\eta _0}} {\int_0^{{{\eta '_0}}} {{p_{AS}}(\eta ){p_{BS}}(\eta ')} } \,\left( {v - \eta m} \right)\,d\eta \,d\eta '\\
\\
b'' = \int_0^{{\eta _0}} {\int_0^{{{\eta '_0}}} {{p_{AS}}(\eta ){p_{BS}}(\eta ')} } \,\left( {v - \eta 'm} \right)\,d\eta \,d\eta '\\
\\
c'' = \int_0^{{\eta _0}} {\int_0^{{{\eta '_0}}} {{p_{AS}}(\eta ){p_{BS}}(\eta ')} } \,\sqrt {\eta \,\eta '} m\,d\eta \,d\eta ' .
\end{array}
\end{eqnarray}\\
Note, in setting  the gains as described above  we must assume that the satellite itself has measured each of the transmittivities separately.  Again, the  final state ensemble at the ground stations is a non-Gaussian mixture.

\section{Comparison of The Schemes}

From the final CM of each scheme the logarithmic negativity ${E_{LN}}$ is adopted  as a measure of the  entanglement between the two ground stations.  As noted above, the resulting ensemble-averaged state shared by the ground stations in each scheme is a non-Gaussian state, and as such cannot be described completely by its first and second moments. Therefore, the entanglement measure we compute based on the CM of the resulting mixed state will represent only the Gaussian entanglement between the terrestrial stations.

We simulate the performance of each of our three schemes in terms of the Gaussian entanglement derived from the appropriate CM.
For all simulations shown calculations of $E_{_{LN}}$ will adopt base 2 in the logarithmic term,    and the following assumptions are adopted:
(i) For each simulation, all initial  states are  two-mode squeezed states with the same initial squeezing $r$.
(ii) Beam wander, as modeled by the log-negative Weibull distribution, is used to characterize the two fading channels for each scheme, with $\beta=1$.
(iii) The two separate fading channels are assumed to be independent, but not necessarily identical.
(iv) The beam wander standard deviations $\,{\sigma _{b\_AS}}\,,\,\,{\sigma _{b\_SA}}\,,\,\,{\sigma _{b\_BS}}\,,\,\,{\sigma _{b\_SB}}$ for the four possible link traversals  satisfy $
{\sigma _{b\_SA}} = {k_1}\,{\sigma _{b\_AS}}\,,\,\,
{\sigma _{b\_BS}} = {k_2}\,{\sigma _{b\_AS}}\,,\,\,{\sigma _{b\_SB}} = {k_1}\,{k_2}\,{\sigma _{b\_AS}}\,,\,\,{\sigma _{b\_AS}} = {\sigma _b}$, where $0 \le {k_1} \le 1$ and ${k_2} \ge 0$. This allows us to parametrize the beam wander dependence on  geometries and communication direction in terms of the three independent parameters, $\sigma_b$, $k_1$ and $k_2$. For clarity the apertures (and beam-spot radii) will be assumed the same at satellite and ground station - we allow the beam wander alone to model different losses  at these devices (when in receive mode).

\begin{figure}[!t]
    \begin{center}
   {\includegraphics[width=2.9 in, height=3.7 in]{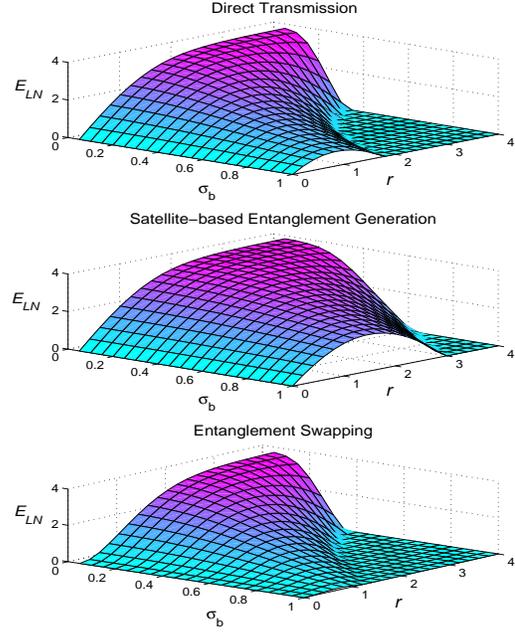}}
    \caption{(Color online) Logarithmic negativity ${E_{LN}}$ of the two-mode state at the ground stations resulting from the direct transmission (top figure), Satellite-based entanglement generation (middle figure) and the entanglement swapping (bottom figure). The results are shown with respect to the beam wander standard deviation ${\sigma _b}$ in the uplink, and the squeezing level $r$. Here,  $\beta/W = 1\,,\,\,{k_1} = 0.5\,,\,\,{k_2} = 0.64$. These parameters for ${\sigma _b}=0.7$ lead to a mean loss of 3dB for the uplink from station A.}\label{fig:1}
    \end{center}
\end{figure}

\begin{figure}[!t]
    \begin{center}
   {\includegraphics[width=2.9 in, height=3.7 in]{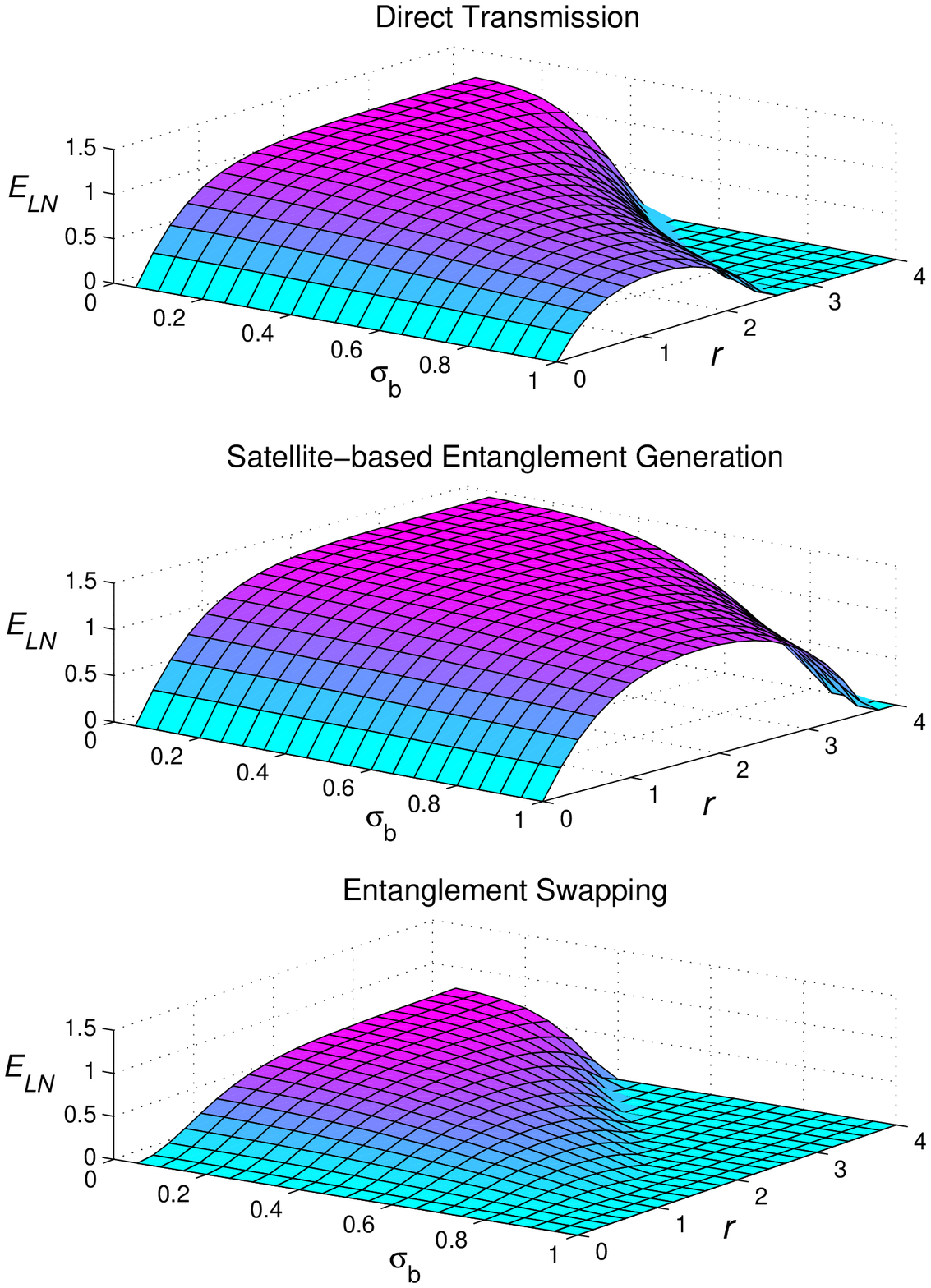}}
    \caption{(Color online) Same as Fig.~\ref{fig:1} except  here  $\beta/W = 0.5$.  These parameters for ${\sigma _b}=0.7$ lead to a mean loss of 5.4dB for the uplink from station A.}\label{fig:2}
    \end{center}
\end{figure}

\begin{figure}[!t]
    \begin{center}
   {\includegraphics[width=2.9 in, height=3.7 in]{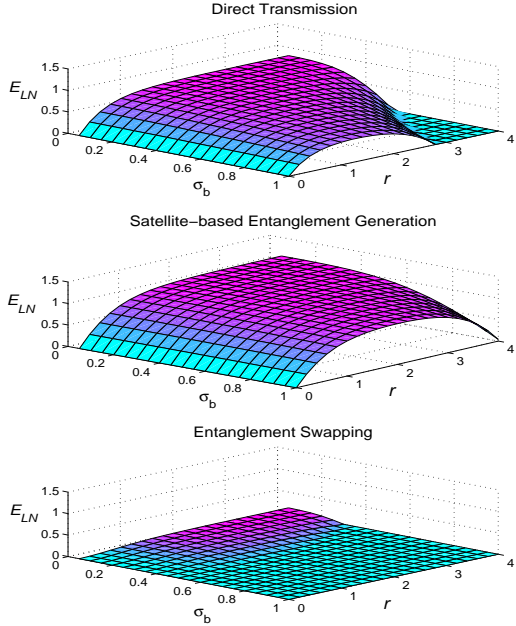}}
    \caption{(Color online) Same as Fig.~\ref{fig:1} except  here, $\beta/W = 0.4$.   These parameters for ${\sigma _b}=0.7$ lead to a mean loss of 6.7dB for the uplink from station A.}\label{fig:3}
    \end{center}
\end{figure}

 Figs.~\ref{fig:1}-\ref{fig:3} show the final Gaussian entanglement of the three communication schemes as a function of beam wander standard deviation ${\sigma_b}$ (normalized to $\beta$) in the uplink from station A, and the squeezing level $r$ of the initial entangled states. The parameters shown  in
 Figs.~\ref{fig:1}-\ref{fig:3} correspond to channels with losses of roughly 4dB through 8dB (at $\sigma_b=1$) in the uplink. They thus represent  low-loss channels such as HAP-LEO satellite channels where the effects of the turbulent atmosphere are relatively small \cite{hap}. Such channels  are also typical of short-length ($\sim$km) atmospheric FSO links as expected at ground level \cite{21}.

Considering each scheme, it is evident that an increase in $\sigma_b$ reduces entanglement while increasing the input squeezing is able to partly compensate its negative effect. For a large squeezing level we see the output logarithmic negativity degrades with increasing squeezing since strongly squeezed states are more sensitive to fading.
However, the main point we wish to draw from these results is that although the satellite-based entanglement generation scheme is always best, its advantage over the direct transmission scheme is rather small in the low-loss channels considered in  Figs.~\ref{fig:1}-\ref{fig:3}. We do note that satellite-based entanglement generation holds a channel advantage in that it does not utilize any uplink. As such, increases in the quality of the downlink channel relative to an uplink channel, beyond that examined here, will lead to a corresponding increase in the entanglement advantage for the satellite-based entanglement scheme relative to the other schemes.

We also note that the direct transmission scheme can always be configured to deliver a better entanglement outcome than the entanglement swapping scheme.
For low values of input squeezing, the swapped Gaussian entanglement between the two terrestrial stations is smaller than the final Gaussian entanglement of the direct transmission. However, for some channel parameters (\emph{e.g.} see Fig.~\ref{fig:1}) there are values in the high-squeezing regime for which the swapping-based scheme can lead to more entanglement than direct transmission. However, for each level of $\sigma_b$, if the optimal amount of squeezing for initial entangled states is used, direct transmission can always be configured to distribute better entanglement than the swapping scheme. Indeed, we can show that for any fading channel,  entanglement swapping can never lead to any improved entanglement generation relative to direct transmission (see Appendix B). These observations on CV entanglement swapping in relation to direct transmission have been noted before for the case of fixed attenuation (non-fading) channels \cite{19}.

Beyond noise terms introduced via atmospheric turbulence, additional losses and noise can occur at the detector including  background light, dark counts, and electronic noise. Losses in the receiver components are typically well below atmospheric losses, and can be accounted for by modification to the estimated transmission factors. Background light can be controlled by sufficient shielding and filtering and is not a serious issue for homodyne detection \cite{R4}, and dark counts are now at the level of less than 20s$^{-1}$ in modern detectors \cite{r10}. As such, electronic noise will dominate receiver noise. Note, any other noise (e.g., that introduced in the channel by an eavesdropper) can be included in this receiver noise so as to form the total \emph{excess} noise.

It is possible to accommodate the introduction of electronic  noise  by the appropriate addition of extra variance  terms ${\chi}$ in each of  our covariance matrices.  In our matrix ${M_{\eta \,\eta '}}$  such a noise term appears on the lower diagonal term; in $M '_{\eta \,\eta '}$ on both diagonal terms, in $M_\eta ^{1,2}$ in the lower diagonal term, and in $M_{\eta '}^{3,4}$  on the upper diagonal term (all additional noise terms multiplied by $I$).

To quantify the effect electronic noise  can have (all results shown thus far have assumed zero excess noise) we have carried out a series of additional simulations where appropriate noise terms have been added for each scheme.  We have assumed the same amount of  noise $\chi$ in all relevant  receivers, and simulated levels of $\chi$ in the range 0.01-0.05, a range consistent with current detectors \cite{R5}. For such a range we find that  the Gaussian entanglement for the direct transmission scheme is  reduced by approximately 2-9 percent, with the entanglement reduction for the other two schemes both approximately 4-16 percent. Such results are consistent with the fact  the simulations of the direct transmission scheme includes noise at only one receiver.

Ground-to-satellite communications are anticipated to undergo much stronger losses than those illustrated in Figs.~\ref{fig:1}-\ref{fig:3}, with single FSO uplink channels anticipated to have losses of order 25dB and beyond \cite{fso,r7}. Under such losses, distribution of  entanglement between the ground stations will be a fruitless endeavor without the intervention of a highly-selective post-selection strategy.
\section{Post-selection}
Here, processing strategies which enhance the Gaussian entanglement of the non-Gaussian mixed state between the ground stations are investigated.  The post-selection strategies which occur at the receiving ground station can be based on classical measurements of the channel transmittance, or on quantum measurements. We are interested in quantifying the performance of these two different measurement strategies. Note that in both post-selection strategies Gaussification occurs in the sense that the conditioned states are more Gaussian in nature due to the enhanced concentration of low-loss states in the final ensemble. For clarity we will study  post-selection strategies in the context of our lowest complexity scheme, namely, the direct transmission scheme.
\subsection{Classical Post-selection}
Although fading noise diminishes Gaussian entanglement, it also provides the possibility to recover it. Post-selection  of large transmission windows, as introduced in \cite{21} for the case of a single fading channel, offers a possibility for improving the Gaussian entanglement in cases where it was strongly diminished by the wider fading.  In this scenario, a subset of the channel transmittance distribution, with high transmittivity, is selected to contribute to the resulting post-selected state.

 For this form of post-selection to operate in our direct transmission scheme, coherent (classical) light pulses are reflected of the satellite in order to measure the transmittance of the combined channel $\zeta  = \eta \,\eta '$ at the receiving ground station, where again $\eta $ and $\eta '$ are random variables describing transmission factors of the uplink and downlink, respectively. The received quantum state is kept or discarded, conditioned on the classical measurement outcome being larger or smaller than the post-selection threshold ${\zeta _{th}}$.
Providing we have a form for the probability density distribution $p(\zeta)$,  the resulting post-selected CM can be calculated as
\begin{eqnarray}\label{PS2}
\begin{array}{l}
{M^{ps}} = \left( {\begin{array}{*{20}{c}}
{\,v\,I}&{\,{c^{ps}}Z}\\
{{c^{ps}}Z}&{\,{b^{ps}}I}
\end{array}} \right) ,

\noindent {{\rm where}} \\
\\
{b^{ps}} = \frac{1}{{{P_s}}}\int_{{\zeta _{th}}}^{{\eta _0}{{\eta '_0}}} {p(\zeta )} \,\left( {1 + \zeta \left( {v - 1} \right)} \right)\,d\zeta \,\\
\\
{c^{ps}} = \frac{1}{{{P_s}}}\int_{{\zeta _{th}}}^{{\eta _0}{{\eta '_0}}} {p(\zeta )} \,\sqrt \zeta  \sqrt {{v^2} - 1} \,\,d\zeta \, .
\end{array}
\end{eqnarray}


Here, ${P_s}$ is the total probability for the combined channel transmission to fall within the post-selected region, and  is given by
\begin{eqnarray}\label{PS3}
{P_s} = \int_{{\zeta _{th}}}^{{\eta _0}{{\eta '_0}}} {p(\zeta )} \,d\zeta \, .
\end{eqnarray}

Using ${M^{ps}}$, the Gaussian entanglement in terms of the logarithmic negativity of the post-selected state can be computed. This is illustrated in Fig.~\ref{fig:PS1}  with respect to the post-selection threshold ${\zeta _{th}}$ and  success probability ${P_s}$, respectively (solid lines). Note that in these calculations no closed-form solution for $p(\zeta)$ could be used, so a numerically determined form was utilized. Fig.~\ref{fig:PS1} explicitly shows for this specific fading channel the trade-off in increased Gaussian entanglement  (as the threshold value increases) at the cost of lower success probability. The other curves (dashed) in this figure relate to quantum post-selection, which is discussed next.
\subsection{Quantum Post-selection}
As we have just seen, classical post-selection offers the possibility of concentrating the Gaussian entanglement at the ground station. However, this comes at additional complexity in the transmission and detection strategy at the ground stations, due to the requirement for ongoing reliable channel estimation. As such, it is useful to explore how Gaussian entanglement concentration may be possible without such channel estimation. To investigate this we will generalize to the combined fading channel, the distillation scheme recently proposed by  \cite{23} for the single fading channel.

Recalling that in the direct transmission scheme, one mode (beam $A$) from the initial two-mode entangled state is at ground station A and the other mode (beam $B$) is transmitted to ground station B via a relaying satellite. From Eq.\eqref{CP4aa}, the CM of the two-mode Gaussian state between the terrestrial stations after each realization of $\eta $ and $\eta '$ can be re-written as
\begin{eqnarray}\label{Dis1}
\begin{array}{l}
{M _{\eta \eta '}} = \left( {\begin{array}{*{20}{c}}
{{v}}&0&{{c_q}}&0\\
0&{{v}}&0&{{c_p}}\\
{{c_q}}&0&{{b_q}}&0\\
0&{{c_p}}&0&{{b_p}}
\end{array}}  \right) , \\
\\
{b_q} = {b_p} = 1 + \eta \,\eta '\left( {v - 1} \right) ,
{c_q} = -{c_p} = \sqrt {\eta \,\eta '} \sqrt {{v^2} - 1} .
\end{array}
\end{eqnarray}
Entanglement distillation is implemented at the receiving ground station by extracting a small portion (beam $t$) of the received mixed state using a tap beam splitter with transmittivity of $T$ and reflectivity of $R=1-T$. A single quadrature (for instance, the amplitude quadrature, ${\hat q_t}$) is then measured on the tapped beam. If the measurement outcome is above the threshold value ${q_{th}}$, then the remaining state (beam $B'$) is kept, otherwise it is discarded. The Wigner function of the state before the beam splitter for each realization of $\eta$ and $\eta '$ is given by
\begin{eqnarray}\label{Dis_A1}
{W_{\eta \eta '}}\left( {{q_A},{p_A},{q_B},{p_B}} \right) = \frac{{\exp \left( { - \frac{1}{2}{R_{AB}}M_{\eta \eta '}^{ - 1}R_{AB}^T} \right)}}{{4{\pi ^2}\sqrt {\det {M_{\eta \eta '}}} }},
\end{eqnarray}
where $R_{AB} = \left( {{q_A},{p_A},{q_B},{p_B}} \right)\,$. Given the Wigner function for the vacuum state as
${W_v}\left( {{q_v},{p_v}} \right) = \frac{1}{{2\pi }}\exp \left( { - \frac{1}{2}\left( {q_v^2 + p_v^2} \right)} \right)
$,
the conditional Wigner function of the output state after distillation for each realization of $\eta$ and $\eta '$ is given by
\begin{eqnarray}\label{Dis_A2}
\begin{array}{l}
W_{\eta \eta '}^d\left( {{q_A},{p_A},{q_{B'}},{p_{B'}}} \right) = \\
\\
\,\,\,\,\,\,\,\int_{{q_{th}}}^\infty  {d{q_t}\int_{ - \infty }^\infty  {d{p_t}\,} } {W_{\eta \eta '}}\left( {{q_A},{p_A},{{\tilde q}_B},{{\tilde p}_B}} \right){W_v}\left( {{{\tilde q}_v},{{\tilde p}_v}} \right),
\end{array}
\end{eqnarray}
where ${\tilde q_B} = \sqrt T {q_{B'}} + \sqrt {R} {q_t}$, ${\tilde p_B} = \sqrt T {p_{B'}} + \sqrt {R} {p_t}$, ${\tilde q_v} = \sqrt T {q_t} - \sqrt {R} {q_{B'}}$, and ${\tilde p_v} = \sqrt T {p_t} - \sqrt {R} {p_{B'}}$.
From the resultant Wigner function, $W_{\eta \eta '}^d$, the moments of the quadrature operators after the distillation for each realization of $\eta$ and $\eta '$ can be calculated [33]. Since $W_{\eta \eta '}^d$ is a Gaussian distribution of the quadrature variables, these moments  can be written as
\begin{eqnarray}\label{Dis4}
\begin{array}{l}
{\left\langle {{q_A}} \right\rangle _{\eta \eta '}} = \frac{{\sqrt R \,{c_q}}}{{\sqrt {2\pi {V_{t,q}}} }}\exp \left( {\frac{{ - q_{th}^2}}{{2{V_{t,q}}}}} \right)\\
\\
{\left\langle {{q_{B'}}} \right\rangle _{\eta \eta '}} = \frac{{\sqrt {TR} \,\left( {{b_q} - 1} \right)}}{{\sqrt {2\pi {V_{t,q}}} }}\exp \left( {\frac{{ - q_{th}^2}}{{2{V_{t,q}}}}} \right)\\
\\
{\left\langle {q_A^2} \right\rangle _{\eta \eta '}} = \frac{{R\,c_q^2{q_{th}}}}{{\sqrt {2\pi V_{t,q}^3} }}\exp \left( {\frac{{ - q_{th}^2}}{{2{V_{t,q}}}}} \right) + \frac{{{v}}}{2}{\rm Erfc} \left( {\frac{{{q_{th}}}}{{\sqrt {2{V_{t,q}}} }}} \right)\\
\\
{\left\langle {q_{B'}^2} \right\rangle _{\eta \eta '}} = \frac{{R\,T{{\left( {{b_q} - 1} \right)}^2}{q_{th}}}}{{\sqrt {2\pi V_{t,q}^3} }}\exp \left( {\frac{{ - q_{th}^2}}{{2{V_{t,q}}}}} \right)\\
\,\,\,\,\,\,\,\,\,\,\,\,\,\,\,\,\,\,\,\,\,\,\,\,\,\,\,\, + \frac{{R\,T{{\left( {{b_q} - 1} \right)}^2} + {b_q}}}{{2{V_{t,q}}}}{\rm Erfc}\left( {\frac{{{q_{th}}}}{{\sqrt {2{V_{t,q}}} }}} \right)\\
\\
{\left\langle {{q_A}{q_{B'}}} \right\rangle _{\eta \eta '}} = \frac{{\sqrt T R\,\left( {{b_q} - 1} \right){c_q}{q_{th}}}}{{\sqrt {2\pi V_{t,q}^3} }}\exp \left( {\frac{{ - q_{th}^2}}{{2{V_{t,q}}}}} \right)\\
\,\,\,\,\,\,\,\,\,\,\,\,\,\,\,\,\,\,\,\,\,\,\,\,\,\,\,\,\,\,\,\,\,\,\,\,\,\,\,\,\,\,\,\,\,\,\,\,\,\,\,\,\,\, + \frac{{\sqrt T {c_q}}}{2}{\rm Erfc} \left( {\frac{{{q_{th}}}}{{\sqrt {2{V_{t,q}}} }}} \right),
\end{array}
\end{eqnarray}
where $\left\langle . \right\rangle $ denotes the expectation value and ${V_{t,q}} = R{b_q} + T$. The elements of the total CM of the resulting distilled state are calculated by averaging over all possible transmission factors of the two fading channels giving the final distilled CM
\begin{eqnarray}\label{Dis2}
\begin{array}{l}
{M ^d} = \left( {\begin{array}{*{20}{c}}
{a_q^d}&0&{c_q^d}&0\\
0&{a_p^d}&0&{c_p^d}\\
{c_q^d}&0&{b_q^d}&0\\
0&{c_p^d}&0&{b_p^d}
\end{array}} \right)  , {\rm where} \\
\\
a_q^d = \left\langle {q_A^2} \right\rangle  - {\left\langle {{q_A}} \right\rangle ^2}\\
\\
b_q^d = \left\langle {q_{B'}^2} \right\rangle  - {\left\langle {{q_{B'}}} \right\rangle ^2}\\
\\
c_q^d = \left\langle {{q_A}} \right\rangle \left\langle {{q_{B'}}} \right\rangle  - \left\langle {{q_A}{q_{B'}}} \right\rangle \\
\\
a_p^d = \frac{1}{P_s}\int_0^{{\eta _0}} {\int_0^{{{\eta '_0}}} {{p_{AS}}(\eta ){p_{SB}}(\eta ')} } {P_{\eta \eta '}}\,{v}\,d\eta \,d\eta '\\
\\
b_p^d = \frac{1}{P_s}\int_0^{{\eta _0}} {\int_0^{{{\eta '_0}}} {{p_{AS}}(\eta ){p_{SB}}(\eta ')} } {P_{\eta \eta '}}\left( {T{b_p} + R} \right)d\eta \,d\eta '\\
\\
c_p^d = \frac{1}{P_s}\int_0^{{\eta _0}} {\int_0^{{{\eta '_0}}} {{p_{AS}}(\eta ){p_{SB}}(\eta ')} } {P_{\eta \eta '}}\,\sqrt T {c_p}\,d\eta , \,d\eta ' \\ \\
\left\langle {{q_A}} \right\rangle  = \frac{1}{{{P_s}}}\int_0^{{\eta _0}} {\int_0^{{{\eta '_0}}} {{p_{AS}}(\eta ){p_{SB}}(\eta ')} } \,{\left\langle {{q_A}} \right\rangle _{\eta \eta '}}\,d\eta \,d\eta '\\
\\
\left\langle {{q_{B'}}} \right\rangle  = \frac{1}{{{P_s}}}\int_0^{{\eta _0}} {\int_0^{{{\eta '_0}}} {{p_{AS}}(\eta ){p_{SB}}(\eta ')} } \,{\left\langle {{q_{B'}}} \right\rangle _{\eta \eta '}}\,d\eta \,d\eta '\\
\\
\left\langle {q_A^2} \right\rangle  = \frac{1}{{{P_s}}}\int_0^{{\eta _0}} {\int_0^{{{\eta '_0}}} {{p_{AS}}(\eta ){p_{SB}}(\eta ')} } \,{\left\langle {q_A^2} \right\rangle _{\eta \eta '}}\,d\eta \,d\eta '\\
\\
\left\langle {q_{B'}^2} \right\rangle  = \frac{1}{{{P_s}}}\int_0^{{\eta _0}} {\int_0^{{{\eta '_0}}} {{p_{AS}}(\eta ){p_{SB}}(\eta ')} } \,{\left\langle {q_{B'}^2} \right\rangle _{\eta \eta '}}\,d\eta \,d\eta '\\
\\
\left\langle {{q_A}{q_{B'}}} \right\rangle  = \frac{1}{{{P_s}}}\int_0^{{\eta _0}} {\int_0^{{{\eta '_0}}} {{p_{AS}}(\eta ){p_{SB}}(\eta ')} } \,{\left\langle {{q_A}{q_{B'}}} \right\rangle _{\eta \eta '}}\,d\eta \,d\eta ' ,
\end{array}
\end{eqnarray}
and where
\begin{eqnarray}\label{Dis3}
\begin{array}{l}
{P_{\eta \eta '}} = \frac{1}{2}{\rm Erfc} \left( {\frac{{{q_{th}}}}{{\sqrt {2{V_{t,q}}} }}} \right) ,\\
\\
{P_s} = \int_0^{{\eta _0}} {\int_0^{{{\eta '_0}}} {{p_{AS}}(\eta ){p_{SB}}(\eta ')} } \,{P_{\eta \eta '}}\,\,d\eta \,d\eta ' .
\end{array}
\end{eqnarray}
Note that here, ${P_s}$ is now the total success probability of distilling the mixed state, and $P_{\eta \eta '}$ is implicitly dependent on $\eta$ and $\eta '$ through the Wigner function $W_{\eta \eta '}^d(q_A, p_A, q_{B'}, p_{B'})$  - but is not to be confused with $p(\zeta)$ defined for the classical post-selection.
\begin{figure}[!t]
    \begin{center}
   {\includegraphics[width=3.4 in, height=2.4 in]{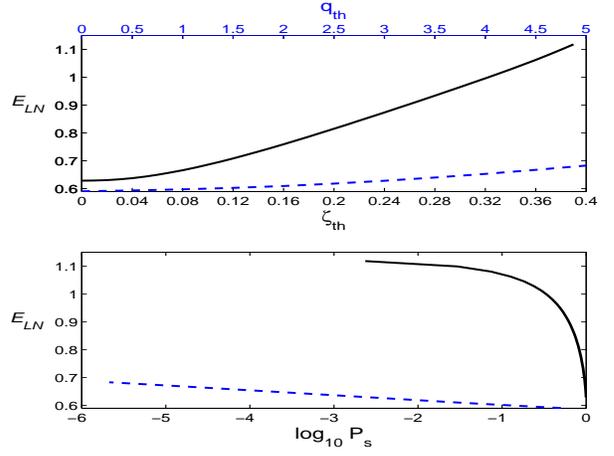}}
    \caption{(Color online) Logarithmic negativity ${E_{LN}}$ of the two-mode states at the ground stations  (for the direct transmission scheme) in terms of the classical post-selection threshold $\zeta _{th}$ (solid line in top figure), quantum post-selection threshold ${q_{th}}$ (dashed line in top figure), and success probability of classical/quantum post-selection ${P_s}$ (bottom figure). Here, $T=0.93$, $r=1.5$, $\beta/W = 0.5\,,\,\,{\sigma _b} = \beta\,,\,\,{k_1} = 0.5\,,\,\,{k_2} = 0.64$. This channel corresponds to a mean loss of 6.4dB in the uplink, and 4.4dB in the downlink.}\label{fig:PS1}
    \end{center}
\end{figure}

\begin{figure}[!t]
    \begin{center}
   {\includegraphics[width=3.4 in, height=2.4 in]{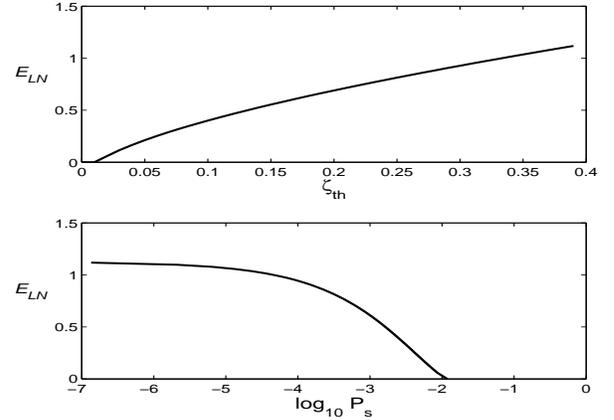}}
    \caption{Logarithmic negativity ${E_{LN}}$ of the two-mode states at the ground stations  (for the direct transmission scheme) in terms of the classical post-selection threshold $\zeta _{th}$ (top figure), and success probability of classical post-selection ${P_s}$ (bottom figure). Here,  $r=1.5$,  $\beta /W = 0.5,\,\,{\sigma _{b\_AS}}\, = 22\beta ,\,\,{\sigma _{b\_SB}}\, = 2\beta $. This channel corresponds to a mean loss of 30dB in the uplink, and 10dB in the downlink.}\label{fig:PS2}
    \end{center}
\end{figure}

Using ${M^{d}}$, the Gaussian entanglement in terms of the logarithmic negativity of the quantum post-selected state can be computed. This is illustrated in Fig.~\ref{fig:PS1}  with respect to the post-selection threshold ${q_{th}}$ and  success probability ${P_s}$, respectively (dashed lines).
Similar to the classical post-selection, we see the amount of Gaussian entanglement is increased by the action of the quantum distillation. However, it is evident that the improvement in Gaussian entanglement is more probable by the classical post-selection. Furthermore, considering the same success probability of each strategy, the classical post-selection is able to generate more Gaussian entanglement compared to the quantum post-selection protocol. Although improvements in the quantum post-selection strategy can be made, due to its direct selection of better channels the classical post-selection scheme will always provide a better result. The results described here illustrate the price to be paid for deploying simpler transmission/detection strategies (no channel estimation) at the ground stations.

Our final result is to look at the entanglement generation rates in the high-loss scenario where the direct transmission scheme is utilized with terrestrial ground stations. In such scenarios one could expect typically 25-30dB loss in the uplink and 5-10dB in the downlink. Fig.~\ref{fig:PS2} shows an example of such a link scenario. Here the quantum post-selection strategy is not shown, as its success probability is found to be too small in such high-loss scenarios.  We can see from Fig.~\ref{fig:PS2} that for the specific channel shown, levels of Gaussian entanglement at $E_{LN}>1$ can be found for success probabilities $<10^{-4}$. These success probabilities can be multiplied by the  transmission rates (currently of order $10^8$Hz) in order to  obtain a mode (pair) generation rate at $E_{LN}>1$ of $10^4$Hz. Note, this should only be considered as a typical rate for the duration of a single pass of a LEO satellite  - as the channel characteristics will vary during the actual LEO pass-over timescale (which is of order a few hundred seconds  \cite{r10}).

\section{Conclusions and Future Directions}
In deploying quantum communications, we are largely faced with three options, the use of fibers, the use of free-space channels, or the use of satellite-based communications. These technologies are complementary and all will likely play a role in the emerging global quantum communication infrastructure. Fiber technology has the key advantage that once in place, an undisturbed channel from A to B exists. However, fiber suffers from large losses which therefore limit its distance - although such distance limitations may be overcome by the development of suitable quantum repeaters \cite{5}. Replacing the fiber channel with a free-space channel has the immediate advantage of fewer losses \cite{r7}, but such a channel is subject to potential ground-dwelling line-of-sight (LoS) blockages, and is also ultimately distance-limited by the visible horizon. Nonetheless, free-space optical communication has a role to play in many scenarios \cite{hap, aircraft}. Free-space quantum communication via satellite has the additional advantage that communication can take place when there is no direct free-space LoS from A to B  in place. Assuming LoS from a satellite to the two ground stations exists, satellite-based communication can proceed. The range of this communication is also potentially much larger than that allowed for by a direct ground-based free-space connection (no terrestrial horizon limit and lower losses at high altitudes). Use of  satellites  also allows for fundamental studies on the impact of  relativity on quantum communications \cite{R2}. The key disadvantage of satellite-based quantum communications is turbulence induced losses, the subject of this work.

In this work we have explored a  range of quantum communication architectures anticipated to play a role in next generation satellite-based communication systems and quantified the expected entanglement generation rates they give rise to. We have focussed on the trade-off between the quantum complexity introduced at the satellite and the resultant Gaussian quantum entanglement between two ground stations (or HAPs). We have found that for low-loss fading channel characteristics  a low-complexity direct transmission scheme (reflection at the satellite) will produce CV entanglement generation rates at the ground stations not too dissimilar from those anticipated for a scheme based on entanglement generated at the satellite itself. For high-loss channels  we find  that a direct transmission scheme can provide for useful levels of entanglement generation. When the downlink channels can be assured to be significantly better than uplink channels, entanglement generation within the satellite will provide for a corresponding significant improvement in entanglement rates at the ground stations - albeit at the cost of embedding quantum systems in the satellite.  In all cases we find entanglement swapping at the satellite to be an inferior solution.

We have also investigated the role played by post-selection in concentrating the entanglement between the ground stations. More specifically, we have investigated the price to be paid if simple transmission and detection strategies are adopted at the ground station in which no channel estimation is required. The quantum post-selection techniques can be utilized in such scenarios, but in general will provide reduced entanglement outcomes relative to classical post-selection techniques based on channel estimation. In high-loss channels classical post-selection is required.

Given the losses anticipated in satellite-based communications, future work should focus on additional effects that lead to enhanced protection of entangled modes transmitted through a turbulent atmosphere. Of particular value would be the use of coding techniques applied to the CV states, use of non-Gaussian states as initial transmission modes, and use of quantum feedback control between the two ground stations. The payoff of such techniques would likely be of most value in a direct transmission scheme. The reliability of CV versions of quantum applications such as quantum key distribution (QKD) \cite{rn1} and quantum location verification \cite{rn2} over high-loss atmospheric fading channels are also worthy of  investigation.  Consideration of hybrid CV/single-photon architectures in the deployment of such techniques and applications would be of particular interest. Finally, we note the application of spatial-diversity techniques as applied to  FSO communications, and the ability of full diversity to be achieved even for transmitters only a few cm apart \cite{mimo}. The role of such diversity techniques in compensating the large losses in uplink ground-satellite channels is also worth exploring in the context of Gaussian entanglement distribution.

\section{Acknowledgments}
This work has been funded by the University of New South Wales (Australia). The authors gratefully acknowledge valuable discussions with Hendra Nurdin.

\appendix

For completeness we detail the analysis that leads to the solutions in our adopted entanglement swapping scheme. Here, we follow closely the fixed-attenuation analysis of \cite{19}, generalizing it to the case of combined fading channels.
\subsection{Entanglement Swapping Covariance Matrices }
Here we wish to highlight how \eqref{21} and \eqref{16s} of the main text are derived.
Let us consider entanglement swapping involving two pairs of entangled modes, one pair consists of modes 1 and 2 and the second pair consists of modes 3 and 4. We assume that the two pairs are described by two Gaussian states, having different CMs and zero first moments, \emph{i.e.},
\begin{eqnarray}\label{1app}
\begin{array}{l}
M _{1,2}^{} = \left( {\begin{array}{*{20}{c}}
{aI}&C\\
{{C^T}}&{bI}
\end{array}} \right),\,C = diag\left ( {{c_ + },{c_ - }} \right )\\
\\
M _{3,4}^{} = \left( {\begin{array}{*{20}{c}}
{dI}&F\\
{{F^T}}&{eI}
\end{array}} \right),\,F = diag\left( {{f_ + },{f_ - }} \right)\\
\\
\left\langle {{{\hat R}_{1,2}}} \right\rangle  = \left\langle {{{\hat R}_{3,4}}} \right\rangle  = 0 .
\end{array}
\end{eqnarray}
In the Wigner function formalism, the initial 4-mode state is described by the product of the Wigner function of two input states
\begin{eqnarray}\label{2}
\begin{array}{l}
{W_{in}}({R_{1,2,3,4}}) = {W_{in}}({R_{1,2}})\,\,{W_{in}}({R_{3,4}})\\
\\
{W_{in}}({R_{i,j}}) \propto \exp ( - \frac{1}{2}{R_{i,j}}\,M _{i,j}^{ - 1}\,R_{i,j}^T) ,
\end{array}
\end{eqnarray}
where $i,j  \in \left\{ {1,2,3,4} \right\}$.
The inverse of CMs, ${M _{1,2}}$ and ${M _{3,4}}$ can be computed as
\begin{eqnarray}\label{3}
\begin{array}{l}
M _{1,2}^{ - 1} = \left( {{\lambda _{ij}}} \right) = \left( {\begin{array}{*{20}{c}}
{\frac{b}{{ab - c_ + ^2}}}&0&{\frac{{ - {c_ + }}}{{ab - c_ + ^2}}}&0\\
0&{\frac{b}{{ab - c_ - ^2}}}&0&{\frac{{ - {c_ - }}}{{ab - c_ - ^2}}}\\
{\frac{{ - {c_ + }}}{{ab - c_ + ^2}}}&0&{\frac{a}{{ab - c_ + ^2}}}&0\\
0&{\frac{{ - {c_ - }}}{{ab - c_ - ^2}}}&0&{\frac{a}{{ab - c_ - ^2}}}
\end{array}} \right)\\
\\
M _{3,4}^{ - 1} = \left( {{\gamma _{ij}}} \right) = \left( {\begin{array}{*{20}{c}}
{\frac{e}{{de - f_ + ^2}}}&0&{\frac{{ - {f_ + }}}{{de - f_ + ^2}}}&0\\
0&{\frac{e}{{de - f_ - ^2}}}&0&{\frac{{ - {f_ - }}}{{de - f_ - ^2}}}\\
{\frac{{ - {f_ + }}}{{de - f_ + ^2}}}&0&{\frac{d}{{de - f_ + ^2}}}&0\\
0&{\frac{{ - {f_ - }}}{{de - f_ - ^2}}}&0&{\frac{d}{{de - f_ - ^2}}}
\end{array}} \right) .
\end{array}
\end{eqnarray}
Thus, the Wigner function of the 4-mode state before the swapping can be given as
\begin{eqnarray}\label{4}
\begin{array}{l}
W_{_{in}}^{}({R_{1,2,3,4}}) \propto \\
\\
\exp \left( { - \frac{1}{2}\left( {{R_{1,2}}\,M _{1,2}^{ - 1}\,R_{1,2}^T + {R_{3,4}}\,M _{3,4}^{ - 1}\,R_{3,4}^T} \right)} \right) = \\
\\
\exp \left\{ { - \frac{1}{2}\left( {{\lambda _{11}}\,q_1^2 + {\lambda _{33}}\,q_2^2} \right.} \right. + {\lambda _{22}}\,p_1^2 + {\lambda _{44}}\,p_2^2\\
\\
{\gamma _{11}}\,q_3^2 + {\gamma _{33}}\,q_4^2 + {\gamma _{22}}\,p_3^2 + {\gamma _{44}}\,p_4^2\\
\\
\left. {\left. { + 2{\lambda _{13}}\,{q_1}{q_2} + 2{\lambda _{24}}\,{p_1}{p_2} + 2{\gamma _{13}}\,{q_3}{q_4} + 2{\gamma _{24}}\,{p_3}{p_4}} \right\}} \right) .
\end{array}
\end{eqnarray}
The swapping is first performed by mixing two modes 2 and 3 through a balanced beam splitter, yielding output modes $u$ and $v$ which at the level of quadrature variables are described by
\begin{eqnarray}\label{5}
\begin{array}{l}
{q_u} = \frac{1}{{\sqrt 2 }}\left( {{q_2} - {q_3}} \right),\,\,\,{p_u} = \frac{1}{{\sqrt 2 }}\left( {{p_2} - {p_3}} \right)\\
\\
{q_v} = \frac{1}{{\sqrt 2 }}\left( {{q_2} + {q_3}} \right),\,\,\,{p_v} = \frac{1}{{\sqrt 2 }}\left( {{p_2} + {p_3}} \right) .
\end{array}
\end{eqnarray}
  With these relations, the Wigner function of the new 4-mode  state after the beam-splitter, ${W_{BS}}({R_{1,4,u,v}})$, can then be obtained from the Wigner function \eqref{4}, namely
\begin{eqnarray}\label{6}
{W_{BS}}({R_{1,4,u,v}}) \propto \exp ( - \frac{1}{2}{R_{1,4,u,v}}\,\,M _{1,4,u,v}^{ - 1}\,\,R_{1,4,u,v}^T) ,
\end{eqnarray}
where
\begin{eqnarray}\label{7}
\begin{array}{l}
{R_{1,4,u,v}} = \left( {{q_1},{p_1},{q_4},{p_4},{q_u},{p_v},{q_v},{p_u}} \right)\\
\\
M _{1,4,u,v}^{ - 1} = \left( {\begin{array}{*{20}{c}}
{{\lambda _{11}}}&0&0&0\\
0&{{\lambda _{22}}}&0&0\\
0&0&{{\gamma _{33}}}&0\\
0&0&0&{{\gamma _{44}}}\\
{k{\lambda _{13}}}&0&{ - k{\gamma _{13}}}&0\\
0&{k{\lambda _{24}}}&0&{k{\gamma _{24}}}\\
{k{\lambda _{13}}}&0&{k{\gamma _{13}}}&0\\
0&{k{\lambda _{24}}}&0&{ - k{\gamma _{24}}}
\end{array}} \right.\\
\\
\left. {\,\,\,\,\,\,\,\,\,\,\,\,\,\,\,\,\,\,\,\,\,\,\,\,\,\,\begin{array}{*{20}{c}}
{k{\lambda _{13}}}&0&{k{\lambda _{13}}}&0\\
0&{k{\lambda _{24}}}&0&{k{\lambda _{24}}}\\
{ - k{\gamma _{13}}}&0&{k{\gamma _{13}}}&0\\
0&{k{\gamma _{24}}}&0&{ - k{\gamma _{24}}}\\
{\frac{1}{2}{\delta _1}}&0&{\frac{1}{2}{\delta _3}}&0\\
0&{\frac{1}{2}{\delta _2}}&0&{\frac{1}{2}{\delta _4}}\\
{\frac{1}{2}{\delta _3}}&0&{\frac{1}{2}{\delta _1}}&0\\
0&{\frac{1}{2}{\delta _4}}&0&{\frac{1}{2}{\delta _2}}
\end{array}} \right) ,
\end{array}
\end{eqnarray}
and where
\begin{eqnarray}\label{8}
\begin{array}{l}
{\delta _1} = {\lambda _{33}} + {\gamma _{11}}\,,\,\,{\delta _2} = {\lambda _{44}} + {\gamma _{22}} ,\\
\\
{\delta _3} = {\lambda _{33}} - {\gamma _{11}}\,,\,\,{\delta _4} = {\lambda _{44}} - {\gamma _{22}}\,,\,\,k = \frac{{\sqrt 2 }}{2} .
\end{array}
\end{eqnarray}
Then, the new quadratures ${\hat q_u}$ and ${\hat p_v}$ are measured with two homodyne detectors, providing the outcomes ${q'_u}$ and ${p'_v}$ with probability $P\left( {{{q'_u}},{{p'_v}}} \right)$. As a result of this measurement, the initial 4-mode state conditionally collapses into a 2-mode state consisting of modes 1 and 4. The Wigner function of this conditional output state  is obtained by integrating ${W_{BS}}({R_{1,4,u,v}})$ over the unmeasured quadratures ${q_v},{p_u}$, giving
\begin{eqnarray}\label{9}
{W_{cond}}({R_{1,4}}) \propto \int {\int {{W_{BS}}({R_{1,4,u,v}})\,d{q_v}\,d{p_u}} } \left| {_{{q_u} = {{q'_u}},\,{p_v} = {{p'_v}}\,}} \right .
\end{eqnarray}
To make progress we use the  partial Gaussian integral formulation for $n$ variables for the case where we wish to integrate over the last $n-m$ of them, \emph{viz}.,
\begin{eqnarray}\label{10}
\begin{array}{l}
\int { \ldots \int {\exp \left[ { - \frac{1}{2}{q^T}Qq} \right]} } \,d{q_{m + 1}}\,\, \ldots \,\,d{q_n}\\
\\
\,\,\,\,\,\,\,\,\,\,\,\,\,\,\,\,\,\,\,\,\,\,\,\,\,\,\,\,\,\,\,\,\,\,\,\,\,\,\,\,\,\, \propto \exp \left\{ { - \frac{1}{2}{u^T}Uu} \right\} ,
\end{array}
\end{eqnarray}
where
\begin{eqnarray}\label{11}
\begin{array}{l}
Q = \left( {\begin{array}{*{20}{c}}
{{U_0}}&V\\
{{V^T}}&{{W_0}}
\end{array}} \right),\,\,U = {U_0} - V\,W_0^{ - 1}\,{V^T}\, ,\\
\\
q = \left( {\begin{array}{*{20}{c}}
u\\
w
\end{array}} \right),\,\,u = \left( {\begin{array}{*{20}{c}}
{{q_1}}\\
 \vdots \\
{{q_m}}
\end{array}} \right),\,\,w = \left( {\begin{array}{*{20}{c}}
{{q_{m + 1}}}\\
 \vdots \\
{{q_n}}
\end{array}} \right) .
\end{array}
\end{eqnarray}
Comparing \eqref{10} with our problem for integrating ${W_{BS}}({R_{1,4,u,v}})$ over the quadratures ${q_v},{p_u}$, we will have
\begin{eqnarray}\label{12}
\begin{array}{l}
{U_0} = \left( {\begin{array}{*{20}{c}}
{{\lambda _{11}}}&0&0&0&{k{\lambda _{13}}}&0\\
0&{{\lambda _{22}}}&0&0&0&{k{\lambda _{24}}}\\
0&0&{{\gamma _{33}}}&0&{ - k{\gamma _{13}}}&0\\
0&0&0&{{\gamma _{44}}}&0&{k{\gamma _{24}}}\\
{k{\lambda _{13}}}&0&{ - k{\gamma _{13}}}&0&{\frac{1}{2}{\delta _1}}&0\\
0&{k{\lambda _{24}}}&0&{k{\gamma _{24}}}&0&{\frac{1}{2}{\delta _2}}
\end{array}} \right)\\
\\
V = \left( {\begin{array}{*{20}{c}}
{k{\lambda _{13}}}&0\\
0&{k{\lambda _{24}}}\\
{k{\gamma _{13}}}&0\\
0&{ - k{\gamma _{24}}}\\
{\frac{1}{2}{\delta _3}}&0\\
0&{\frac{1}{2}{\delta _4}}
\end{array}} \right),\,\,{W_0} = \left( {\begin{array}{*{20}{c}}
{\frac{1}{2}{\delta _1}}&0\\
0&{\frac{1}{2}{\delta _2}}
\end{array}} \right)\\
\\
q = {R_{1,4,u,v}}\,,\,\,\,u = \left( {\begin{array}{*{20}{c}}
{{q_1}}\\
{{p_1}}\\
{{q_4}}\\
{{p_4}}\\
{{q_u}}\\
{{p_v}}
\end{array}} \right),\,\,\,w = \left( {\begin{array}{*{20}{c}}
{{q_v}}\\
{{p_u}}
\end{array}} \right) .
\end{array}
\end{eqnarray}
The Wigner function for the conditional state of modes 1 and 4 is then given by
\begin{eqnarray}\label{13}
\begin{array}{l}
W_{_{cond}}^{}({R_{1,4}}) \propto \exp \left\{ { - \frac{1}{2}{u^T}Uu} \right\}\\
\\
U = \left( {\begin{array}{*{20}{c}}
{{U_{11}}}&0&{{U_{13}}}&0&{{U_{15}}}&0\\
0&{{U_{22}}}&0&{{U_{24}}}&0&{{U_{26}}}\\
{{U_{13}}}&0&{{U_{33}}}&0&{{U_{35}}}&0\\
0&{{U_{24}}}&0&{{U_{44}}}&0&{{U_{46}}}\\
{{U_{15}}}&0&{{U_{35}}}&0&{{U_{55}}}&0\\
0&{{U_{26}}}&0&{{U_{46}}}&0&{{U_{66}}}
\end{array}} \right)
\end{array}
\end{eqnarray}
where
\begin{eqnarray}\label{14}
\begin{array}{l}
{U_{11}} = \frac{{e(b + d) - f_ + ^2}}{{a(de - f_ + ^2) + e(ab - c_ + ^2)}},{U_{13}} = \frac{{ - {c_ + }{f_ + }}}{{a(de - f_ + ^2) + e(ab - c_ + ^2)}}\\
\\
\,{U_{15}} = \frac{{ - \sqrt 2 e{c_ + }}}{{a(de - f_ + ^2) + e(ab - c_ + ^2)}},{U_{22}} = \frac{{e(b + d) - f_ - ^2}}{{a(de - f_ - ^2) + e(ab - c_ - ^2)}}\\
\\
\,{U_{24}} = \frac{{{c_ - }{f_ - }}}{{a(de - f_ - ^2) + e(ab - c_ - ^2)}},{U_{26}} = \frac{{ - \sqrt 2 e{c_ - }}}{{a(de - f_ - ^2) + e(ab - c_ - ^2)}}\\
\\
{U_{33}} = \frac{{a(b + d) - c_ + ^2}}{{a(de - f_ + ^2) + e(ab - c_ + ^2)}},{U_{35}} = \frac{{\sqrt 2 a{f_ + }}}{{a(de - f_ + ^2) + e(ab - c_ + ^2)}}\\
\\
{U_{44}} = \frac{{a(b + d) - c_ - ^2}}{{a(de - f_ - ^2) + e(ab - c_ - ^2)}},{U_{46}} = \frac{{ - \sqrt 2 a{f_ - }}}{{a(de - f_ - ^2) + e(ab - c_ - ^2)}}\\
\\
\,{U_{55}} = \frac{{2ae}}{{a(de - f_ + ^2) + e(ab - c_ + ^2)}},{U_{66}} = \frac{{2ae}}{{a(de - f_ - ^2) + e(ab - c_ - ^2)}} .
\end{array}
\end{eqnarray}
From the above we can see that the first moments of the conditional output state depends on the measurement results ${q'_u}$ and ${p'_v}$ and  each of the four quadratures will be proportional to
 ${\left({\begin{array}{*{20}{c}}
{-{{q'_u}}\,{U_{15}}} ,
{ - {{p'_v}}\,{U_{26}}} ,
{ - {{q'_u}}\,{U_{35}}},
{ - {{p'_v}}\,{U_{46}}}
\end{array}}\right)^{T}}$.
Let us consider ${U'}$ as
\begin{eqnarray}\label{15}
U' = \left( {\begin{array}{*{20}{c}}
{{U_{11}}}&0&{{U_{13}}}&0\\
0&{{U_{22}}}&0&{{U_{24}}}\\
{{U_{13}}}&0&{{U_{33}}}&0\\
0&{{U_{24}}}&0&{{U_{44}}}
\end{array}} \right) .
\end{eqnarray}
Thus, the CM of the conditional state of modes 1 and 4 can be obtained by inverting $U'$ to give
\begin{eqnarray}\label{16app}
M _{_{1,4}}^{} = \left( {\begin{array}{*{20}{c}}
{a - \frac{{c_ + ^2}}{{b + d}}}&0&{\frac{{{c_ + }{f_ + }}}{{b + d}}}&0\\
0&{a - \frac{{c_ - ^2}}{{b + d}}}&0&{ - \frac{{{c_ - }{f_ - }}}{{b + d}}}\\
{\frac{{{c_ + }{f_ + }}}{{b + d}}}&0&{e - \frac{{f_ + ^2}}{{b + d}}}&0\\
0&{ - \frac{{{c_ - }{f_ - }}}{{b + d}}}&0&{e - \frac{{f_ - ^2}}{{b + d}}}
\end{array}} \right) .
\end{eqnarray}
By setting the matrices of \eqref{1} to those of \eqref{1app}, we find \eqref{16app} leads to \eqref{16s}. However, the protocol is not complete. A final subtlety is that as it stands this matrix represents the outcome for a one-shot Bell measurement. We still have to average over all Bell measurement results. But as we now show if we optimize our choice of gains in the displacement procedure of the protocol, we will arrive at \eqref{16s} as the final CM averaged over all Bell measurement results (for a specific channel realization).

In order to complete the swapping process, the measurement results are broadcast so that modes 1 and 4 can properly be displaced according to the measurement outcomes ${q'_u}$ and ${p'_v}$. In practice, the displacements should be weighted by gain factors so as to improve the quality of the swapped entanglement. In terms of the quadrature operators, these conditional displacements can be expressed as
\begin{eqnarray}\label{17}
\left\{ \begin{array}{l}
{{\hat q}_1} \to {{\hat q}_1} - {g_1}\sqrt 2 {{q'_u}}\\
{{\hat p}_1} \to {{\hat p}_1} + {g_1}\sqrt 2 {{p'_v}}
\end{array} \right.\,,\,\,\left\{ \begin{array}{l}
{{\hat q}_4} \to {{\hat q}_4} + {g_4}\sqrt 2 {{q'_u}}\\
{{\hat p}_4} \to {{\hat p}_4} + {g_4}\sqrt 2 {{p'_v}}
\end{array} \right.
\end{eqnarray}
where ${g_1}$ and ${g_4}$ are the gain factors for the displacement of modes 1 and 4, respectively.

Using \eqref{17} the first moments of the four quadratures of the displaced conditional state $W_{dis}(R_{1,4})$ are then proportional to
\begin{eqnarray}\label{18}
  \sqrt 2 \left( {\begin{array}{*{20}{c}}
{{{q'_u}}\frac{{ - {g_1}\left( {e(b + d) - f_ + ^2} \right) - {g_4}{c_ + }{f_ + } + e{c_ + }}}{{a(de - f_ + ^2) + e(ab - c_ + ^2)}}}\\
{{{p'_v}}\frac{{{g_1}\left( {e(b + d) - f_ - ^2} \right) + {g_4}{c_ - }{f_ - } + e{c_ - }}}{{a(de - f_ - ^2) + e(ab - c_ - ^2)}}}\\
{{{q'_u}}\frac{{{g_4}\left( {a(b + d) - c_ + ^2} \right) + {g_1}{c_ + }{f_ + } - a{f_ + }}}{{a(de - f_ + ^2) + e(ab - c_ + ^2)}}}\\
{{{p'_v}}\frac{{{g_4}\left( {a(b + d) - c_ - ^2} \right) + {g_1}{c_ - }{f_ - } + a{f_ - }}}{{a(de - f_ - ^2) + e(ab - c_ - ^2)}}}
\end{array}} \right) .
\end{eqnarray}
The Wigner function of the output state   averaged over all possible Bell measurements is therefore given by
\begin{eqnarray}\label{19}
{W_{ens}}({R_{1,4}}) = \int {\int {P({{q'_u}},{{p'_v}}\,){W_{dis}}({R_{1,4}})\,d{{q'_u}}\,d{{p'_v}}\,} } ,
\end{eqnarray}
where $P({{q'_u}},{{p'_v}}\,)$ is the probability density of the Bell measurement outcomes.
This average leads to a zero-mean two-mode Gaussian state with the following CM
\begin{eqnarray}\label{20}
\begin{array}{l}
{M _{ens}} = \left( {\begin{array}{*{20}{c}}
{{m _{11}}}&0&{{m _{13}}}&0\\
0&{{m _{22}}}&0&{{m _{24}}}\\
{{m _{13}}}&0&{{m _{33}}}&0\\
0&{{m _{24}}}&0&{{m _{44}}}
\end{array}} \right)\\
\\
{m _{11}} = a + (b + d)g_1^2 - 2{c_ + }{g_1}\\
\\
{m _{22}} = a + (b + d)g_1^2 + 2{c_ - }{g_1}\\
\\
{m _{33}} = e + (b + d)g_4^2 - 2{f_ + }{g_4}\\
\\
{m _{44}} = e + (b + d)g_4^2 + 2{f_ - }{g_4}\\
\\
{m _{13}} = {c_ + }{g_4} + {f_ + }{g_1} - {g_1}{g_4}(b + d)\\
\\
{m _{24}} = {c_ - }{g_4} + {f_ - }{g_1} + {g_1}{g_4}(b + d) .
\end{array}
\end{eqnarray}
The optimal   choice of  gains are those for which all terms of \eqref{18} equal zero. In such a case the CM of the  averaged state \eqref{20} is equal to that of the conditional state in \eqref{16app}. Assuming phase-independent gains, this optimal point is obtained for ${c_ + } =  - {c_ - } = :c$ and ${f_ + } =  - {f_ - } = :f$, and
\begin{eqnarray}\label{21app}
{g_1} = \frac{c}{{b + d}}\,\,,\,\,{g_4} = \frac{f}{{b + d}} ,
\end{eqnarray}
and the CM of \eqref{16app} is obtained. Again by inspecting matrices  \eqref{1} and \eqref{1app}, we find \eqref{21app} leads to \eqref{21}.

\subsection{Effective Loss Channels}
As shown by \cite{Lund}, any CM of the standard form
\begin{eqnarray}\label{ef1}
M_s  = \left( {\begin{array}{*{20}{c}}
{a\,I}&{c\,Z}\\
{c\,Z}&{b\,I}
\end{array}} \right)
\end{eqnarray}
($a,b,c\in \mathbb{R}$)
which satisfies the uncertainty principle and is entangled (the positive partial transpose (PPT) criterion for separability is violated \cite{Simon}), is equivalent to the CM of a lossy two-mode squeezed state with effective squeezing ${r_{e}}$ and effective channel transmissions $\eta _{e}^a$ and $\eta _{e}^b$ for the first and second modes, respectively. These effective parameters are given by
\begin{eqnarray}\label{ef2}
\begin{array}{l}
\cosh (2{r_{e}}) = \frac{{{c^2} + \left( {a - 1} \right)\left( {b - 1} \right)}}{{{c^2} - \left( {a - 1} \right)\left( {b - 1} \right)}} ,\\
\\
\eta _{e}^a = \frac{{a - 1}}{{\cosh (2{r_{e}}) - 1}}\ \ \ ,
\eta _{e}^b = \frac{{b - 1}}{{\cosh (2{r_{e}}) - 1}} .
\end{array}
\end{eqnarray}
Therefore, the CM for each realization of ${\eta}$ and $\eta '$ which are given by \eqref{CP4aa},  \eqref{sa1} and \eqref{16s} for direct transmission, satellite-based entanglement generation and swapping, respectively, can all be re-written in the context of lossy two-mode squeezed states (of course the first two schemes can be directly seen as loss channels).
Averaging over all possible values of ${\eta}$ and $\eta '$, total effective transmittivities and total effective squeezing can be computed for all three schemes as follows.\\
\break
 \emph{(i) Direct transmission:} \\
\begin{eqnarray}\label{ef4}
\begin{array}{l}
\cosh (2r) = v\,\,,\,\,\,{\eta ^a} = 1\\
\\
{\eta ^b} = \int_0^{{\eta _0}} {\int_0^{{{\eta '_0}}} {{p_{AS}}(\eta )\,{p_{SB}}(\eta ')} } \,\eta \,\eta '\,d\eta \,d\eta ' .
\end{array}
\end{eqnarray}\\
\emph{(ii) Satellite-based entanglement generation:} \\
\begin{eqnarray}\label{ef4-1}
\begin{array}{l}
\cosh (2r') = v\,\\
\\
{\eta ^{a'}} = \int_0^{{\eta _0}} {{p_{SA}}(\eta )} \,\eta \,d\eta \,\,\,,\,\,{\eta ^{b'}} = \int_0^{{{\eta '_0}}} {{p_{SB}}(\eta )} \,\eta '\,d\eta ' .
\end{array}
\end{eqnarray}\\
\emph{(iii) Entanglement swapping:}\\
\begin{eqnarray}\label{ef4-2}
\begin{array}{l}
\cosh (2r'') = \int_0^{{\eta _0}} {\int_0^{{{\eta '_0}}} {{p_{AS}}(\eta )\,{p_{BS}}(\eta ')} } \cosh (2\,{{r''_{\eta \,\eta '}}})d\eta \,d\eta '\\
\\
\cosh (2{{r''_{\eta \,\eta '}}}) = \frac{{\left( {{\eta ^2} + {{\eta '}^2}} \right)\left( {1 - v} \right) + \eta \eta '\left( {{v^2} + 3} \right) + \left( {\eta  + \eta '} \right)\left( {v - 3} \right) + 2}}{{\left( {\eta  + \eta ' - 1} \right)\left( {\left( {\eta  + \eta '} \right)\left( {v - 1} \right) + 2} \right)}}\\
\\
{\eta ^{a''}} = \int_0^{{\eta _0}} {\int_0^{{{\eta '_0}}} {{p_{AS}}(\eta )\,{p_{BS}}(\eta ')} } \,\frac{{ - \left( {\eta  + \eta ' - 1} \right)\left( {v - 1} \right)}}{{\left( {\eta \left( {1 - v} \right) + 2\left( {\eta ' - 1} \right)} \right)}}d\eta \,d\eta '\\
\\
{\eta ^{b''}} = \int_0^{{\eta _0}} {\int_0^{{{\eta '_0}}} {{p_{AS}}(\eta )\,{p_{BS}}(\eta ')} } \,\,\frac{{ - \left( {\eta  + \eta ' - 1} \right)\left( {v - 1} \right)}}{{\left( {\eta '\left( {1 - v} \right) + 2\left( {\eta  - 1} \right)} \right)}}d\eta \,d\eta ' .
\end{array}
\end{eqnarray}
\\
 Given the following constraints;   $1 < v < \infty $, $1 < \cosh (2r'') < \infty $, and $0 \le {\eta ^a} \le 1$ (likewise ${\eta ^b},{\eta ^{a'}},{\eta ^{b'}},{\eta ^{a''}},{\eta ^{b''}}$)
it is straightforward to show that
 the total effective transmittivity ${\eta ^{a''}}{\eta ^{b''}}$ for the swapping scheme is always less than or equal to  ${\eta ^a}{\eta ^b}$ for direct transmission.  In addition, we know that in practice, the overall effective transmittivity ${\eta ^{a'}}{\eta ^{b'}}$ for the satellite-based entanglement generation is larger than the overall transmittivity ${\eta ^a}{\eta ^b}$ for the direct transmission since the mean value of the transmittance for the fading downlink is always larger than that for the fading uplink.
  Therefore, the result that the order of the best performance is  (1) satellite-based entanglement generation, (2) direct transmission, (3) entanglement swapping, is a result that in practice will hold.


\begin{thebibliography}{1}
\bibitem{foot1} We do appreciate that in all satellite-based systems ``simple" is very much a relative term. Command and control (both classical)  of reflectors, in relation to tracking and point-ahead angles,  are  non-trivial engineering tasks, but tasks largely removed from direct quantum communication and quantum control.
\bibitem{foot2} Here we assume transceiver gains and other system parameters are set so as to largely compensate free-space divergence losses, and that other atmospheric losses are small in comparison to beam wander losses. Optical system losses will be assumed negligible, or be accounted for by an additional constant attenuation factor. We will also assume receivers possess classical-feedback control systems (adaptive optics) that minimize the beam wander.
\end{thebibliography}
\end{document}